\documentclass[12pt, preprint]{emulateapj}

\usepackage{apjfonts}
\usepackage{multirow}
\usepackage{color}
\usepackage{hyperref}
\bibpunct[; ]{(}{)}{;}{a}{}{;}

\newcommand{\beq}{\begin{equation}}
\newcommand{\eeq}{\end{equation}}
\newcommand{\beqa}{\begin{eqnarray}}
\newcommand{\eeqa}{\end{eqnarray}}

\def\boom{{\sc BOOM\-ER\-anG }}

\shorttitle{\boom Measurement of SZ Fluctuations}
\shortauthors{Veneziani et al.}

\begin{document}
\title{Subdegree Sunyaev-Zel'dovich Signal from Multifrequency \boom observations}
\author{
M. Veneziani,\altaffilmark{1,2,3}
A. Amblard\altaffilmark{1},
A. Cooray\altaffilmark{1},
F. Piacentini\altaffilmark{2},
D. Pietrobon\altaffilmark{4,5},
P. Serra\altaffilmark{1},
P.~A.~R. Ade\altaffilmark{6},
J.~J. Bock\altaffilmark{7,8},
J.~R. Bond\altaffilmark{9},
J. Borrill\altaffilmark{10},
A. Boscaleri\altaffilmark{11},
P. Cabella\altaffilmark{4},
C.~R. Contaldi\altaffilmark{12},
B.~P. Crill\altaffilmark{7,8},
P. de Bernardis\altaffilmark{2},
G. De Gasperis\altaffilmark{4},
A. de Oliveira-Costa\altaffilmark{13},
G. De Troia\altaffilmark{4},
G. Di Stefano\altaffilmark{14},
K.~M. Ganga\altaffilmark{3},
E. Hivon\altaffilmark{15},
W.~C. Jones\altaffilmark{16},
T.~S. Kisner\altaffilmark{17},
A.~E. Lange\altaffilmark{7},
C.~J. MacTavish\altaffilmark{18},
S. Masi\altaffilmark{2},
P.~D. Mauskopf\altaffilmark{6},
A. Melchiorri\altaffilmark{2},
T.~E. Montroy\altaffilmark{17},
P. Natoli\altaffilmark{4},
C.~B. Netterfield\altaffilmark{19},
E. Pascale\altaffilmark{19},
G. Polenta\altaffilmark{2,20,21},
S. Ricciardi\altaffilmark{2,22},
G. Romeo\altaffilmark{14},
J.~E. Ruhl\altaffilmark{17},
P. Santini\altaffilmark{2},
M. Tegmark\altaffilmark{13},
N. Vittorio\altaffilmark{4}
}

\email{marcella.veneziani@roma1.infn.it, amblard@uci.edu}

\altaffiltext{1}{Center for Cosmology, University of California, Irvine, CA 92697, USA}
\altaffiltext{2}{Dipartimento di Fisica, Universit\`a di Roma ``La Sapienza'', Rome, Italy}
\altaffiltext{3}{APC, Universit\'e Paris Diderot, 75013 Paris, France}
\altaffiltext{4}{Dipartimento di Fisica, Universit\`a di Roma ``Tor Vergata'', Rome, Italy}
\altaffiltext{5}{Institute of Cosmology and Gravitation, University of Portsmouth, UK}
\altaffiltext{6}{Department of Physics and Astronomy, Cardiff University, Cardiff, UK}
\altaffiltext{7}{Jet Propulsion Laboratory, Pasadena, CA 91109, USA}
\altaffiltext{8}{California Institute of Technology, Pasadena, CA 91125, USA}
\altaffiltext{9}{CITA, University of Toronto, Toronto, ON M5S 3H8, Canada}
\altaffiltext{10}{Computational Research Division, LBNL, Berkeley, CA 94720, USA}
\altaffiltext{11}{IFAC-CNR, 50127, Firenze, Italy}
\altaffiltext{12}{Theoretical Physics Group, Imperial College, London, UK}
\altaffiltext{13}{Department of Physics, MIT, Cambridge, MA 02139, USA}
\altaffiltext{14}{Istituto Nazionale di Geofisica e Vulcanologia, 00143 Rome, Italy}
\altaffiltext{15}{Institut d\rq Astrophysique de Paris, 75014 Paris, France}
\altaffiltext{16}{Department of Physics, Princeton University, Princeton, NJ 08544, USA}
\altaffiltext{17}{Case Western Reserve University, Cleveland, OH 44106, USA}
\altaffiltext{18}{Astrophysics Group, Imperial College, London, UK}
\altaffiltext{19}{Physics Department, University of Toronto, Toronto ON, Canada}
\altaffiltext{20}{ASI Science Data Center, c/o ESRIN, 00044 Frascati, Italy}
\altaffiltext{21}{INAF - Osservatorio Astronomico di Roma, Monte Porzio Catone, Italy}
\altaffiltext{22}{Space Sciences Laboratory, UC Berkeley CA, USA}


\begin{abstract}

The Sunyaev--Zel'dovich (SZ) effect is the inverse Compton-scattering of cosmic microwave background (CMB) photons by 
hot electrons in the intervening gas throughout the universe. The effect has a distinct spectral signature
that allows its separation from other signals in multifrequency CMB datasets. Using CMB anisotropies measured at
three frequencies by the \boom 2003 flight we constrain SZ fluctuations in the 10 arcmin to 1 deg angular range.
Propagating errors and potential systematic effects through simulations, 
we obtain an overall upper limit of 15.3 $\mu$K (2$\sigma$) for rms SZ fluctuations
in a broad bin between multipoles of 250 and 1200 at the Rayleigh-Jeans (RJ) end of the spectrum.
The resulting upper limit on the local universe normalization of the density perturbations
with \boom SZ data alone is $\sigma_8^{\rm SZ} < 1.14$ at the 95\% confidence level.
When combined with other CMB anisotropy and SZ measurements, we find $\sigma_8^{\rm SZ} < 0.92$ (95\% c.l.).


\end{abstract}

\section{Introduction}
\label{sec:introduction}

The Sunyaev -- Zel'dovich (SZ) effect \citep{SunZel} is scattering of cosmic microwave background (CMB) photons by electrons in the intervening gas throughout the universe. The SZ effect is generally subdivided into two subcomponents: the kinetic effect due to the bulk motion and the thermal effect due to energy transfer from hot electrons. 
The latter is expected to be the largest modification to the background temperature during photon transit from 
the last scattering surface. We will concentrate on it hereafter and refer to it as the SZ signal.
The strongest SZ perturbations to the CMB temperature
are dominated by scattering of photons via hot electrons in massive galaxy clusters \citep{KomKit,Springel,Cooray,Molnar,Seljak,Sadeh}, but
there could also be an SZ signal from reionization \citep{Oh}.
The integrated SZ angular power spectrum is now a known probe of the amplitude of density perturbations, $\sigma_8$ \citep{Komatsu,Bond}.

The SZ effect has been clearly imaged toward individual galaxy clusters \citep{Greggo,Carl,Jones} and has been used for a variety of
applications, including a measurement of the CMB temperature at the redshifts of Coma and A2163 \citep{Battistelli}.
However, the amplitude of the SZ power spectrum at arcminute angular scales, generated from unresolved galaxy clusters, is still not well established with differences at the 2$\sigma$ level from a variety of detections and limits \citep{CBI,ACBAR,BIMA,SZA,QUAD,BOLOCAM}.
Existing SZ anisotropy measurements are restricted to observations with a narrow frequency coverage and to small
areas on the sky. The differences could be a combination of foreground contamination 
and large non-Gaussian variance of the SZ signal~\citep{Cooray01}. Also, no constraints on the SZ signal exist at tens of arcminute 
scales where primary CMB fluctuations dominate.

A clean separation of the SZ anisotropies from primordial CMB is possible due to the fact that the SZ signal 
has a distinct frequency spectrum from the 2.7K blackbody spectrum \citep{Cooray2}. The spectral 
difference arises as inverse-Compton scattering leads to, on average, a 
net energy gain for the CMB photons and the scattered 
photons move from the low frequency Rayleigh--Jeans (RJ) tail to high frequencies \citep{SunZel}.
The SZ sky is colder than the CMB at low frequencies and
hotter than the CMB at high frequencies with no difference at about a
frequency of 217 GHz. A potential detection of the
large-scale structure SZ fluctuations is then aided by observations across
the SZ null from the negative side to the positive side.

A dataset of the form needed for a study of the large-scale structure
SZ effect is provided by the 2003 flight of the balloon-borne
\boom experiment \citep{Masi}. This instrument derives directly from
the \boom payload that was flown in 1998 and resulted in first high
signal-to-noise maps of the CMB anisotropy with sub-horizon
resolution \citep{Bernardis}. The instrument was launched by NASA on
2003 January 6 from Williams Field near McMurdo Station, in
Antarctica. The flight lasted a total of 311 hours until 2003 January 21
and 119 hours of this observing period were devoted to scanning a deep
survey region. The remaining time was spent on scanning a larger
shallow survey and a section of the Galactic plane. Here, we concentrate on a
search for SZ effect in the central deep field over 100 deg$^2$ 
with the highest signal-to-noise ratio.

In this Letter, we report the first statistical limits of the SZ signal 
at sub-degree angular scales at these wavelengths. The discussion is organized as follows: in Section \ref{sec:sepsz} we detail our approach to extract the SZ signal from the multifrequency \boom dataset; 
in Section \ref{sec:3} we detail our simulations used to estimate statistical and systematic uncertainties; 
and in Section \ref{sec:4} we present our results.

\begin{figure}[]
\begin{center}
\includegraphics[width=8cm]{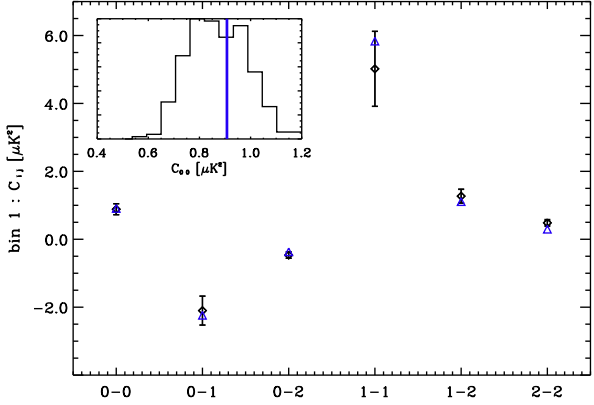}
\end{center}
\caption{Elements of the matrix $C_{ij}$ (Eq.~1) for the data (blue triangle) and simulations
in ($\mu$K)$^2$ RJ units in the first multipole bin ($250<\ell<450$). The six independent points of the $C_{ij}$ matrix correspond
to correlation between frequencies labeled 0, 1, and 2 for 145 GHz, 245 GHz, and 345 GHz respectively.
We show two sets of distributions for simulations. The smaller distribution takes into account only instrumental noise and cosmic variance. The larger distribution includes also primary CMB and foregrounds by taking the rms of three different amplitudes for the components (see text).
In the case where only one error bar is visible, the rms from foreground is smaller than the instrumental noise and cosmic variance.
The inserted plot shows the distribution of $C_{00}$ with the data value inserted as a vertical blue line.}
\label{fig:1}
\end{figure}

\section{CMB and Foreground Removal}
\label{sec:sepsz}

To separate the SZ signal from all other sources of anisotropies, we adopted a
technique well-known in the literature for removing foregrounds from
the CMB anisotropies \citep{Tegmark,Tegmark3,Amblard}. In our case, instead of
recovering the primordial CMB signal, we recover SZ fluctuations by
minimizing the covariance relative to the SZ frequency dependence, and
treat primordial fluctuations as another source of noise. In the remainder of this Letter, we will refer to ``foregrounds'' as all the additional emissions (CMB, Galactic dust, far-IR sources or FIRB, radio point sources) to the SZ effect.

The power spectrum of the SZ can be obtained from a weighted mean of the 
power spectra at different frequencies: $C_l^{\rm SZ} = {\bf w}^T {\bf C} {\bf w}$ \citep{Cooray2}.
The weights $w_l^i$ at each frequency $i$ and multipole $\ell$ can be obtained
by minimizing the covariance of data multipole moments ${\bf C}\equiv \langle \hat{a}^i_{lm} \hat{a}^j_{lm}
\rangle$ subject to the constraint that SZ estimation is unbiased ($\sum_i w(\nu_i)=1$). 

In the case of \boom maps, each frequency channel consists of several
detectors. We distinguish individual detectors with the indices $(\alpha,\beta)$, while $(i,j)$ are indices for the frequency channels.
In order to minimize instrumental noise more aggressively, we compute the covariance matrix of the signal by averaging 
all the combinations of cross-spectra between different detectors and ignoring the auto-spectra of the same detector.
 The contribution of the correlated noise between two different detectors is taken into account in the simulations
and removed as part of a residual contribution to the SZ signal.

We construct the binned covariance matrix in multipole $\ell$ bin $b$ as
\begin{equation}
{\bf C}_{ij} = \sum_{l \in b,m} \sum_{\alpha,\beta} \frac{\langle a_{lm}^{i,\alpha} {a_{lm}^{j,\beta}}^\star \rangle}{s(\nu_i) s(\nu_j) b_l^{i,\alpha}b_l^{j,\beta}} \quad {\rm with} \; \alpha \neq \beta, \; {\rm if} \; i=j \, ,
\end{equation}
where $s(\nu_i)$ is the SZ frequency dependence at each of the
\boom frequency bands relative to CMB with $s(\nu)=2-(x/2) \coth
(x/2)$, $x=h\nu/kT_{\rm CMB}\approx \nu/56.8$GHz, and $b_l^{i,\alpha}$ is the
measured beam window function for the detector $\alpha$ in channel $i$. Note that with the definition 
above, in the RJ limit $s(\nu)\rightarrow 1$ so that $C_l^{\rm SZ}(\nu,\nu') =
s(\nu)s(\nu')C_l^{\rm SZ}$ where $C_l^{\rm SZ}$ is the SZ anisotropy
power spectrum in the RJ limit. 
The covariance matrix C$_{ij}$ is required to be invertible and positive definite. We numerically check this
both in data and simulations. In Figure~\ref{fig:1} we show the covariance matrix from data and 
compare it to simulations described below.

Using the data covariance matrix, the optimal weights for the SZ reconstruction are 
\begin{equation}
{\bf w} = \frac{{\bf C}^{-1} {\bf e}}{{\bf e}^T {\bf C}^{-1} {\bf e}}\, 
\end{equation}

\noindent where ${\bf e}$ is a unit vector, $e(\nu_i)=1$.
The \boom channels consist of eight polarization-sensitive bolometers at 145 GHz, 
and four spider-web bolometers at each of 245 GHz and 345 GHz channels.
We make use of data from all these
detectors except two detectors that were known in prior studies to be
dominated by detector noise (245X and 345Z) \citep{Masi} and two detectors
with a significantly higher noise than the others (145Z2 and 345Y), 
leaving us with seven detectors at 145, three at 245 and two at 345 GHz. 

We use the spectral response of each band as measured in lab with sub-percent accuracy.
From these bands we derived the values of (0.49722, -0.21646, -1.01643) for $s(\nu)$ at 145, 245 and 345~GHz respectively.
These bands provide ideal frequency coverage for an SZ study with channels in the SZ decrement, near the null, and
the increment, respectively. 
The measured FWHM of the beams is 11.5, 8.5, and 9.1~arcmin for the 145, 245, and 345 GHz 
channel, respectively~\citep{Jones2}. These values include a 2.4~arcmin pointing jitter. 
The beams window functions $b_l^{i,\alpha}$ are in fact numerically derived from physical optics simulations, 
combined with a Gaussian pointing jitter. 

Similar to prior studies with \boom data, we 
produced CMB temperature anisotropy (T) maps using the Italian analysis pipeline \citep{Masi} and the TT power spectra
with the MASTER method~\citep{MASTER}. To remove excess atmospheric 
noise, we filtered out time-ordered data in each of the detectors in frequency space below 200~mHz. 
This results in a damping of power at angular scales above 1.2$^\circ$.
The effect of this filter, of the scanning strategy, of the sky coverage, and of the pixelization 
results in an effective window function. This window function is estimated by projecting signal-only sky simulation into 
detector time-streams. To do that we used 10 sky simulations of the SZ signal, 
obtained from \citet{White} with different normalizations and initial conditions. 
These time-streams are then analyzed with our pipeline and the resulting angular 
power spectrum is compared to the power spectrum of the input sky signal 
to obtain the window function. The 10 different simulations result in slightly different window functions. 
The scatter in the window function is included in the final error estimate of the SZ signal. 

\begin{table}[!t]
\caption{\label{tab:1}}
\begin{center}
{\sc SZ Power Spectrum Estimates}
\begin{tabular}{lccc}
\hline
& Bin 1 & Bin 2 & Bin 3\\
$\ell$-range & 250 -- 450& 450 -- 700 & 700 -- 1200 \\
\hline \hline
\multicolumn{4}{c}{Optimal weights}\\
\hline
$w_{\rm 145 GHz}$ & 0.9323 & 0.8514 & 0.7289 \\
$w_{\rm 245 GHz}$ & 0.4193 & 0.3771 & 0.3002 \\
$w_{\rm 345 GHz}$ & -0.3515 & -0.2285 & -0.0292 \\
\hline \hline
Raw SZ & 236 & 164 & 538 \\
\hline \hline
\multicolumn{4}{c}{Residuals $^{(a)}$}\\
\hline
CMB & 53 & 36 & 70\\
Instr. noise & 92 & 12 & -95\\
Galactic dust & 68 & 82 & 138\\
FIRB &44 & 81& 195\\
Radio sources & 3 & 7 & 58\\
\hline
Total residual & 247& 202 & 338\\
\hline \hline
\multicolumn{4}{c}{SZ Band Power Uncertainties $^{(b)}$}\\
\hline
Instr. noise &154 & 116& 280\\
Foregrounds & 37 & 79 & 145\\
Beam & 3 & 5 & 44\\
Calibration & 121& 77 & 63\\
Transfer func. & 2 & 3 & 11\\
Cosmic \& NG Var $^{(c)}$ & 7 & 6 & 4 \\ 
\hline\hline
Final SZ Band Power &-11 $\pm$ 199 & -38 $\pm$ 160 & 200 $\pm$ 325\\
\hline 
\end{tabular}
\end{center}
\footnotesize
NOTES.---%
The weights and Raw SZ designate the weight vectors for each multipole bin and the
SZ power spectrum respectively with both as measured from data.
Except in the case of weights {\bf w},
the values are tabulated in units of $\mu$K$^2$ for the SZ angular power spectrum $l^2C_l/2\pi$ at the RJ end of the frequency spectrum.\\
a: The residuals are the average spectra measured on our SZ-free simulations and represent our bias. 
The total residual is different from the sum of the partial residuals due to small ($<$10\%) random
correlation between components.\\
b: The uncertainties are the dispersion measured with our simulations. 
The final SZ spectrum values are corrected for the noise and foreground bias with the dispersion 
error from simulations. \\
c: Assuming the WMAP team's SZ power spectrum with $\sigma_8=0.95$, the 2$\sigma$ upper limit we derived from all SZ data. 
The calculation for the Non-Gaussian (NG) covariance makes use of the same halo model as used for this power spectrum.

\end{table}

\section{Simulations}\label{sec:3}

Monte Carlo simulations were used in order to estimate the residual signal from correlated detector noise,
primary CMB, and foregrounds that is detected as an SZ signal at the end.
In fact, when we minimize the covariance in the data to extract the SZ, the optimal combination
does not remove perfectly the other signals. 
This is particularly true with just three channels, and is expected to improve with
more channels. We also use simulations to estimate final uncertainties in the SZ signal.

\begin{figure}[]
\begin{center}
\includegraphics[width=8.5cm]{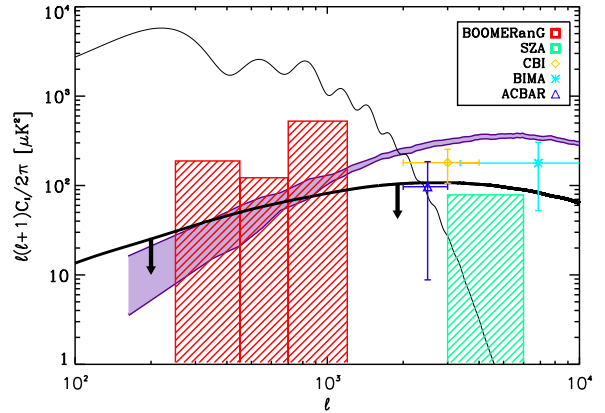}
\end{center}
\caption{Angular power spectrum of SZ anisotropies at the RJ end of the frequency spectrum. We show the 68\% confidence upper limits from the \boom data in hashed columns to the left and
the reported upper limit at the 65\% confidence level from the SZA experiment to the right. The data points from CBI, BIMA, and ACBAR experiments
are also shown (see text for details). We correct the ACBAR point to RJ end of the frequencies. The lines show two theory predictions for SZ fluctuations: 
the curved region is a prediction based on results from numerical simulations with $\sigma_8^{\rm SZ}=1$ and the solid line is the same model as used by the WMAP team but scaled to $\sigma_8^{\rm SZ}=0.95$, the 2-$\sigma$ upper limit from all SZ data. The limit from \boom SZ data  alone is $\sigma_8^{\rm SZ} < 1.14$ (95\% c.l.). }
\label{fig:2}
\end{figure}

We generate 200 time-stream simulations for each one of the 12 detectors with a combination of sky signal simulations that were 
projected into a time-line using the \boom pointing, and a random noise realization with variance consistent with data.
As the covariance matrix C$_{ij}$ of Eq.~1 is built using the cross-spectra between different detectors,
the noise realizations include also correlated noise between different detectors.
For 145 GHz the noise correlation spectra are reported in Table 7 of \citet{Masi} and they are below 3\% for the detectors used
in the analysis.
For the 245 and 345 GHz we measured the noise correlation spectra removing the optimal signal map from the time-ordered data. 
The correlations are at most 6\% for the detectors used in the analysis.

In addition to 200 realizations of primary CMB, 
the sky signal simulations are computed with Galactic dust, far-Infrared background (FIRB) sources, and radio point sources
and combined with \boom bands passes.  The Galactic dust for the observed field is described with model 8 from \citet{Finkbeiner}, while
FIRB and point sources were obtained from the Planck sky model \citep[PSM;][J. Delabrouille 2009, in preparation]{Leach}.

The amplitude of each component in the data was estimated with a Monte Carlo Markov Chain for the
three frequencies together. We compared the measured temperature angular power spectrum with a linear combination 
of CMB, dust, FIRB, and point sources and fitted the amplitude coefficients $p_\gamma$ of each component $\gamma$ 
such that $C_l^{\rm tot}=\sum_\gamma p_\gamma C_l^{\gamma} + N_l$, where $N_l$ is detector noise.
The CMB amplitude of the predicted model is the same as we find in the data with $p_{\rm CMB}=1.01 \pm 0.04$. 
We find $p_{\rm dust}=4.6 \pm 0.8$, larger than predicted by model 8 of \citet{Finkbeiner} and $p_{\rm radio}=0.76 \pm 0.07$ for
 radio point sources, smaller than predicted by \citet{dezotti} counts used in PSM, but in agreement with \citet{QUAD}. 
The amplitude of the FIRB remains unconstrained with $0<p_{\rm FIRB}<0.7$ ($1\sigma$).

To take the uncertainties in the amplitudes of foregrounds into account, we ran three sets of 200 simulations. In the first set we leave 
the level of foregrounds as the models predict ($p_\alpha=1$ for each of CMB, dust, FIRB and 
radio sources). In the second case we set $p_{\rm dust}=4.6$, $p_{\rm FIRB}=0$, and $p_{\rm radio}=0.76$.
In the third case, we change $p_{\rm FIRB}=0.7$, while keeping the rest of the parameters as in the second case.
We note that these simulations do not include the SZ signal as we are reconstructing the SZ under
the assumption of a zero signal. This does not bias the procedure since the optimal weights from Eq.~1 are independent of the exact amplitude of the large-scale SZ effect.

The simulated time-lines have been analyzed with the same pipeline as used for the data.
This results in three sets of Monte-Carlo binned power spectra for the SZ $C_{b\; {\rm MC}}^{\rm SZ}$, which 
should, in principle, be zero since the SZ signal is not included in the simulations. 
From the distribution of each set of $ C_{b\; {\rm MC}}^{\rm SZ}$ we derived
(1) a bias in the SZ binned power spectrum from $\langle C_{b\; {\rm MC}}^{\rm SZ} \rangle$
and (2) the bin-to-bin covariance matrix due to statistical noise and sampling variance of the CMB,
\begin{equation}\label{eqn:pippo}
\mathcal C_{bb'} = \langle \left( C_{b\; {\rm MC}}^{\rm SZ} - \langle C_{b\; {\rm MC}}^{\rm SZ} \rangle\right)\times \left(C_{b'\; {\rm MC}}^{\rm SZ} - \langle C_{b'\; {\rm MC}}^{\rm SZ} \rangle\right) \rangle \, .
\end{equation}
The error bars on the angular power spectrum are given by 
$\Delta C_b^{\rm SZ} = \sqrt{\mathcal C_{bb}}$.
In Table~\ref{tab:1}, we quote the average value of the bias from the three sets in our residual amplitude and add the average of the dispersion of these residuals as an additional error (Table~\ref{tab:1} foreground error).

These foreground errors are combined with the beam errors, which are estimated again through 
Monte Carlo simulations. The calibration uncertainties of the time-lines were also included through Monte Carlo simulations.
They are at most $2\%,\;8\%$, and $13\%$ for the 145 GHz, 245 GHz, and 345 GHz, respectively, leading to 
an error of $4\%,\;16\%$ and $26\%$ on the temperature angular power spectra at each of the three frequencies.
All the uncertainties listed above are added in quadrature for the final SZ band-power uncertainty.

\section{SZ Power Spectrum Estimate}\label{sec:4}

In Table~\ref{tab:1} we list the values we obtain for the three multipole bins between $\ell=$250 and 1200.
We also list the residual level from each foreground component, including detector noise,
and the error associated with various uncertainties as described above. As tabulated, the biggest contamination to SZ detection comes from
instrumental noise at the largest angular scales, while the FIRB dominates the contamination at the smallest angular scales probed by the
experiment. Radio point sources generate negligible confusion, primarily because at these high frequencies radio sources produce a weaker background compared to the dusty galaxies making the FIRB. 
In accounting for foreground contamination in the SZ estimate, we have taken a conservative approach here allowing for all components. An
aggressive approach with the assumption of no FIRB leads to a marginal detection of an SZ signal especially in the third bin.
Though the amplitude of FIRB fluctuations at 350 GHz is uncertain and we have based our model on the PSM, we do not consider an SZ detection with 
a no FIRB assumption to be realistic. 

The binned SZ power spectrum limits at the RJ end of the frequency spectrum are shown in Figure~\ref{fig:2}, where we plot the 68\% confidence level
limit for three bins between multipoles of 250 and 1200. In estimating the final SZ band power uncertainty,
we also include the usual Gaussian cosmic variance and the extra covariance from the non-Gaussian
nature of the SZ power spectrum \citep{Cooray01}. This covariance is calculated assuming $\sigma_8^{\rm SZ}=0.95$ and making use of the
same halo model as the one used by the WMAP team's SZ model and shown with a solid line in Fig.~2 \citep{Komatsu}.
Within this model, we study the cosmological implications of our limit on the SZ fluctuations, using a MCMC package \citep{COSMOMC} to 
constrain the amplitude of fluctuations. For reference to numerical simulations, with a dashed line, we also show the
average SZ signal and the scatter from a set of ten simulations at $\sigma_8=1$ from \citet{White}. 

In addition to three bins shown in Figure~\ref{fig:2}, we also combine the estimation of \boom SZ power spectrum to a single broad bin of
$250 < \ell < 1200$.  We find an upper limit of 234 $\mu$K$^2$ in $l(l+1)C_l/2\pi$ at the 95\% confidence level. 
Previous analytical calculations have shown 
that $C_l^{\rm SZ}\propto (\sigma_8^{\rm SZ})^7 (\Omega_bh)^2$ \citep{Seljak}, where we separate $\sigma_8^{\rm SZ}$ associated with SZ from the primordial normalization $\sigma_8$.  The amplitude constraint from \boom SZ data alone is $\sigma_8^{\rm SZ} < 1.14$ at 95\% confidence level. 

In Figure~\ref{fig:2}, we also compare our upper limits with results on SZ fluctuations in the literature,
including CBI \citep{CBI}, BIMA \citep{BIMA}, and ACBAR \citep{ACBAR}. We scale the ACBAR value from 150 GHz to the RJ end of the spectrum for 
easy comparison with all other results.  We also fit jointly
 the combined WMAP 5-year \citep{WMAP}, ACBAR \citep{ACBAR}, and CBI \citep{CBI} data together with SZ
upper limits from \boom and SZA. We use the same analytical halo model with $\sigma_8^{\rm SZ}=0.95$
to include an extra uncertainty associated with non-Gaussian covariance in each of these measurements; these, however, make only a minor
difference except in the case of BIMA where the smaller area surveyed increase the importance of non-Gaussianities.
Marginalizing over all other cosmological parameters in the $\Lambda$CDM model, 
we find $\sigma_8^{\rm SZ} < 0.92$ at 95\% confidence level ($\sigma_8^{\rm SZ} < 0.71$ at 1$\sigma$).
This SZ derived amplitude is fully consistent with WMAP 5-year result with $\sigma_8=0.81 \pm 0.02$ \citep{WMAP}.
While it has been claimed in the past that the SZ derived $\sigma_8$ is higher than the value derived 
from the CMB, we do not find this is the case with the \boom data. 

The next opportunities to perform a multi frequency analysis similar to ours will be with Planck
and OLIMPO \citep{masi08}. Both these experiments include multiple bands at high frequencies where the SZ is positive.
As we have found that only one channel above the SZ null frequency is not adequate to separate both CMB and FIRB from SZ fluctuations,
with several high frequency channels, these upcoming CMB experiments should be able to measure and 
separate FIRB more accurately than we were able to with just one
channel at 350 GHz.

We gratefully acknowledge support from NSF CAREER AST-0645427 at UCI.
We also acknowledge support from the Italian Space Agency (contracts I/087/06/0 and I/016/07/0), and from Programma Nazionale 
Ricerche in Antartide. The authors acknowledge the use of the Planck Sky Model, 
developed by the Component Separation Working Group (WG2) of the Planck Collaboration.


\begin{thebibliography}{36}
\expandafter\ifx\csname natexlab\endcsname\relax\def\natexlab#1{#1}\fi

\bibitem[{Amblard et al. (2007)}]{Amblard}
Amblard, A., Cooray, A., Kaplinghat, M., \ 2007, PRD, 75, 083508

\bibitem[{Battistelli et al. (2002)}]{Battistelli}
Battistelli, E.~S., et al.,\ 2002, ApJL, 580, L101

\bibitem[{Bond et al. (2005)}]{Bond}
  Bond, J. R. et al.\ 2005, ApJ, 626, 12

\bibitem[{Carlstrom et al. (2002)}]{Carl}
Carlstrom, J. E., Holder, G. P., \& Reese, E. D.\ 2002, ARA\&A, 40, 643

\bibitem[{Cooray(2000)}]{Cooray}
Cooray, A., \ 2000, Phys. Rev. D, 62, 103506

\bibitem[{Cooray(2001)}]{Cooray01}
Cooray, A., \ 2001, Phys. Rev. D, 64, 063514

\bibitem[{Cooray et al. (2000)}]{Cooray2}
Cooray, A., Hu, W. \& Tegmark, M., \ 2000, ApJ, 540, 1

\bibitem[{Cooray \& Sheth(2002)}]{CooShe}
Cooray, A. \& Sheth, R. 2002, Phys. Rep., 372, 1

\bibitem[{Dawson et al. (2006)}]{BIMA}
Dawson, K.~S., et al., \ 2006, ApJ, 647, 13

\bibitem[{de Bernardis et al. (2000)}]{Bernardis} 
de Bernardis, P. et al., \ 2000, Nature, 404, 955

\bibitem[{de Zotti et al. (2005)}]{dezotti}
 de Zotti, G., et al. 2005, A\&A, 431, 893

\bibitem[{Finkbeiner et al. (1999)}]{Finkbeiner} 
Finkbeiner, D. P., Davis, M. \& Schlegel, D. J. \ 1999, ApJ, 524, 867

\bibitem[{Friedman et al. (2009)}]{QUAD}
Friedman, R. B. et al., \ 2009, ApJ, 700, L107

\bibitem[{Greggo et al. (2000)}]{Greggo}
Greggo, L. et al.,\ 2000, ApJ, 539, 39

\bibitem[{Hivon et al. (2002)}]{MASTER}
Hivon, E. et al., \ 2002, ApJ, 567, 2

\bibitem[{Jones et al. (1993)}]{Jones} 
Jones, M. et al.,\ 1993, Nature, 365, 320

\bibitem[{Jones et al. (2006)}]{Jones2} 
Jones, W. C. et al., \ 2006, ApJ, 647, 823

\bibitem[{Komatsu \& Kitayama(1999)}]{KomKit}
Komatsu, E. \& Kitayama, T.\ 1999, ApJ, 526, L1

\bibitem[{Komatsu et al. (2002)}]{Komatsu} 
Komatsu, E. \& Seljak, U., \ 2002, MNRAS, 336, 1256

\bibitem[{Komatsu et al. (2008)}]{WMAP} 
Komatsu, E. et al., \ 2009, ApJS, 180, 330

\bibitem[{Leach et al. (2008)}]{Leach}
Leach, S. M., et al. 2008, A\&A, 491, 597

\bibitem[{Lewis \& Bridle(2002)}]{COSMOMC}
Lewis, A. \& Bridle, S., \ 2002, PRD, 66, 103511

\bibitem[{Masi et al. (2006)}]{Masi} 
Masi, S. et al., \ 2006, A\&A, 458, 687

\bibitem[{Masi et al. (2008)}]{masi08} 
Masi, S. et al.,\ 2008, Mem. Soc. Astron. Ital., 79, 887

\bibitem[{Molnar \& Birkinshaw(2000)}]{Molnar}
  Molnar, S. \& Birkinshaw, M. 2000, ApJ, 537, 542

\bibitem[{Oh et al. (2003)}]{Oh} 
Oh, S.-P., Cooray, A. \& Kamionkowski, \ 2003, MNRAS, 342, L20

\bibitem[{Reichardt et al. (2008)}]{ACBAR} 
Reichardt, C.~L. et al., \ 2008, ApJ, 694, 1200

\bibitem[{Sadeh \& Rephaeli(2004)}]{Sadeh}
  Sadeh, S. \& Rephaeli, Y. 2004, NA, 9, 373

\bibitem[{Sayers et al. (2008)}]{BOLOCAM}
Sayers, J. et al., \ 2008, ApJ, 690, 1597

\bibitem[{Seljak et al. (2001)}]{Seljak}
Seljak, U., Burwell, J. \& Pen, Ue-Li, \ 2001, Phys. Rev. D, 63, 063001

\bibitem[{{Sharp} {et al.} (2009)}]{SZA}
Sharp, M. K. et al.,\ 2009, ApJ, submitted (arXiv:0901.4342)

\bibitem[{Sievers et al. (2009)}]{CBI}
Sievers, J. L. et al., \ 2009, arXiv:0901.4540

\bibitem[{Springel et al. (2001)}]{Springel} 
Springel, V., White, M. \& Hernquist, L., \ 2001, ApJ, 549, 681

\bibitem[{Sunyaev et al. (1972)}]{SunZel} 
Sunyaev, R.~A. \& Zeldovich, Ya.~B.,\ 1972, Comments Astrophys. Space Phys., 4, 173

\bibitem[{Tegmark et al. (2003)}]{Tegmark3}
Tegmark, M., de Oliveira-Costa A., Hamilton A., \ 2003, Phys. Rev. D, 68, 123523

\bibitem[{Tegmark \& Efstathiou(1996)}]{Tegmark}
Tegmark, M. \& Efstathiou, G., \ 1996, MNRAS, 281, 1297

\bibitem[{White(2003)}]{White} 
White, M., \ 2003, ApJ, 597, 650


\end{thebibliography}
\end{document}